\documentclass[epsfig,aps,twocolumn]{revtex4}
\usepackage{amssymb}
\usepackage{amsmath}
\usepackage{graphicx}
\usepackage{lscape}
\usepackage{booktabs}

\begin{document}

\title{Forward and backward comparative study of jet properties in $pp$ collisions at $\sqrt{s}$=7 TeV}

\author{Yu-Liang Yan$^{1,2}$ \footnote{yanyl@ciae.ac.cn}, Ayut Limphirat$^{3}$ \footnote{ayut@g.sut.ac.th}, Dai-Mei Zhou$^2$,
Pornrad Srisawad$^{4}$, Yupeng Yan$^{3}$, Chunbin Yang$^2$, Xu Cai$^2$, and Ben-Hao Sa$^{1,2}$ \footnote{sabh@ciae.ac.cn}}

\affiliation{$^1$ China Institute of Atomic Energy, P. O. Box 275 (10),
              Beijing, 102413 China. \\
             $^2$ Key Laboratory of Quark and Lepton Physics (MOE) and
              Institute of Particle Physics, Central China Normal University,
              Wuhan 430079, China.\\
             $^3$ School of Physics and Center of Excellence in High Energy Physics $\& $ Astrophysics, Suranaree University of
Technology, Nakhon Ratchasima 30000, Thailand.\\
             $^4$ Department of Physics, Naresuan University, Phitsanulok 65000, Thailand.}

\begin{abstract}
We propose a forward method, based on the PYTHIA6.4, to study theoretically the jet
properties in the ultra-relativistic $pp$ collisions. In the forward method, the
partonic initial states are first generated with PYTHIA6.4 and then hadronized in
the Lund srting fragmentation regime, and finally hadronic jets are constructed
from the created hadrons. Jet properties calculated in the forward method for
$pp$ collisions at $\sqrt{s}$=7 TeV are comparable with the corresponding
ones calculated with usual anti-$k_t$ algorithm (backward method) in the PYTHIA6.4.
The comparison between the results in the backward and forward methods may
bring benefit to the understanding of the partonic origin of jets in the backward method.
\end{abstract}

\maketitle

\section {Introduction}
In the early stage of ultra-relativistic heavy ion collisions, the hard parton
scatterings generate high transverse momentum partons  which traverse the
medium and then hadronize into sprays of particles called jets \cite{Connors}.
Jet studies play an important role in understanding the properties of the
medium created in ultra-relativistic heavy-ion collisions \cite{Solana}. The
``jet-quenching" together with the ``elliptic flow" reveal much essential
characters of the strongly coupled quark-gluon plasma (sQGP) in the
ultra-relativistic heavy ion collisions at RHIC \cite{rhic1,rhic2,rhic3,rhic4}
and LHC \cite{ALICE1,CMS1,ATLAS1} energies. With the higher collision energy,
higher luminosity, and better detectors for jet measurements, the LHC
experiments are able to measure the jet properties more precisely. Recently,
the ATLAS, CMS, and ALICE collaborations have published a series results of
jet properties in the pp, p+Pb, and Pb+Pb collisions at LHC energies
\cite{ATLAS2,ATLAS3,ATLAS4,ATLAS5,ATLAS6,ALICE2,CMS2,CMS3}, where some new
physics is arisen and needs to be studied urgently.

The perturbative Quantum Chromodynamics (pQCD) is able to describe the hard
parton-parton scattering process and related initial and final state parton
showers quantitatively. However, one can not apply the pQCD to directly
describe the properties of
particles within the jets, as the hadronization of partons is a
non-perturbative process. To predict the intra-jet properties,
phenomenological models may have to be employed. Several Monte Carlo (MC)
event generators, like PYTHIA6.4 \cite{PYTHIA6.4}, PYTHIA8 \cite{PYTHIA8.2},
HERWIG \cite{HERWIG}, HERWIG++ \cite{HERWIG++}, and SHERPA \cite{SHERPA} etc.,
are available in the market. The results of MC event generators are sensitive
to the model parameters assumed, which must be tuned to fit experimental data.
In the end a number of PYTHIA tunes, such as Perugia 2011 tune \cite{Perugia} etc.,
exist.

In the PYTHIA model, the Leading Order perturbative QCD (LO-pQCD) is used to
generate the $2\rightarrow 2$ hard processes (parton-parton collisions).
The parton shower model \cite{ps1,ps2,ps3} is employed to describe the initial
and final state parton radiations. As for the soft processes (non-perturbative
processes) several empirical options are available\cite{under1,under3}.
Finally, the Lund string hadronization regime is used
for the hadronization of partons \cite{string1,string2,string3}. In this work,
the PYTHIA6.4 \cite{PYTHIA6.4} is employed to investigate the properties of
particles in the jets in $pp$ collisions at $\sqrt{s}$=7 TeV using both the
backward (usual anti-$k_t$ algorithm) and forward methods.

\section {Backward method for jet study}
\begin{figure}[htbp]
\includegraphics[width=0.45\textwidth]{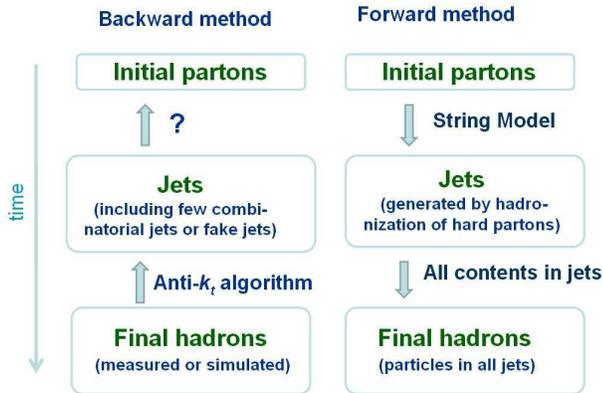}
\caption{(color online) A sketch of backward method vs. forward method.}
\label{jet}
\end{figure}

In the backward method we first employ PYTHIA6.4 to generate the final
hadronic state for pp collisions at $\sqrt{s}$=7 TeV. Then the jet-finding
algorithms, i.e. the anti-$k_t$ technique \cite{antik,Salam}, is used to
backward reconstruct the jets as done in usual experimental jet analysis
\cite{ATLAS3}. From the final hadronic state to the reconstructed hadronic
jets, and to the searching for their partonic origin, this routine is opposite
to the nature time evolution process of collisions, as sketched in
the left part of Fig.(\ref{jet}). Hence it is referred to as
backward method.

In the anti-$k_t$ algorithm, the distance $d_{ij}$ between entities (particle
or energetic cluster) $i$ and $j$ is defined as
\cite{antik}:
\begin{equation}
d_{ij}={\rm min}(k^{-2}_{ti},k^{-2}_{tj})\frac{(\Delta R)^2_{ij}}{R^2},
\end{equation}
where
\begin{equation}
(\Delta R)^2_{ij}=(y_i-y_j)^2+(\phi_i-\phi_j)^2,
\end{equation}
and $k_{ti}$, $y_i$ and $\phi_i$ are the transverse momentum, rapidity and
azimuthal angle of particle $i$, respectively. The distance between particles
$i$ and beam (B) $d_{iB}$ is defined as
\begin{equation}
d_{iB}=k^{-2}_{ti}.
\end{equation}

With the distances $d_{ij}$ and $d_{iB}$, a list is compiled containing all
the $d_{ij}$ and $d_{iB}$ for the particles in an event. If the smallest
entry is a $d_{ij}$, particles $i$ and $j$ are combined (their four-vectors
are added) as a jet. If the smallest entry is a $d_{iB}$, the particle $i$ is
considered as a complete jet and removed from the list. The distances for all
entities are recalculated and the procedure repeated until no entities are
left. Thus the distance parameter $R$ in anti-$k_t$ algorithm is an essential
quantity, within which the particles are reconstructed to a jet. If a hard
particle has no hard neighbor within the distance 2$R$, then it will simply
accumulate all the soft particles within a circle of radius $R$, resulting in
a perfectly conical jet. If there is another hard particle $2$ in the area of
$R < \Delta R_{12} < 2R$, then there will be two jets. The shape of jet $1$
will be conical and jet $2$ will be partly conical when $k_{t1} \gg k_{t2}$;
or both cones will be clipped when $k_{t1} \sim k_{t2}$ \cite{antik}. The
anti-$k_t$ algorithm is infrared and collinear safe and produces geometrically
``conelike" jets, so it is widely used for jet reconstruction in experimental
data analysis \cite{ATLAS2,ALICE2,CMS2}. The jets are reconstructed from the
final hadronic state and hopefully to be attributed to the initial partonic
state.

Based on the hadronic final states of $pp$ collisions at $\sqrt{s}$=7 TeV
generated by the PYTHIA6.4, we use the anti-$k_t$ algorithm to reconstruct the
jets. Following ATLAS \cite{ATLAS3} the charged particles with transverse
momentum $p_T >$ 300 MeV/c and pseudorapidity $|\eta|<2.5$ are counted and
the distance parameter $R$=0.6 is used. After the jet reconstruction, the
clusters with $p_T >$ 4 GeV/c and $|y|<$ 1.9 are accepted as the jets,
including those with only one particle in the cluster. The jets are divided
into five bins according to their transverse momentum $p_{\rm{T,jet}}$, namely
4-6, 6-10, 10-15, 15-24 and 24-40 GeV/c. The jets with $p_T >$ 40 GeV/c and
the particles not belong to any jet are excluded in the calculations.

After jets have been reconstructed in the anti-$k_t$ algorithm, we calculate the
intra-jet particle distributions \cite{ATLAS3}:
\begin{eqnarray}\label{distributions}
F(z, p_{\rm{T,jet}})\equiv\frac{1}{N_{jet}}\frac{dN_{ch}}{dz}, \nonumber \\
f(p_T^{rel}, p_{\rm{T,jet}})\equiv\frac{1}{N_{jet}}\frac{dN_{ch}}{dp_T^{rel}}, \nonumber \\
\rho_{ch}(r, p_{\rm{T,jet}})\equiv\frac{1}{N_{jet}}\frac{dN_{ch}}{2\pi rdr}.
\end{eqnarray}
The $N_{ch}$ and $N_{jet}$ in Eq. (\ref{distributions}) are respectively
the charged particle and jet numbers in a given jet $p_{\rm{T,jet}}$ bin.
The variable $z$ (known as the fragmentation variable)
\begin{equation}
z=\frac{\vec{p}_{ch}\cdot \vec{p}_{jet}}{|\vec{p}_{jet}|^2}
\label{eqz}
\end{equation}
is defined for each charged particle in a jet, and
the variable $r$ stands for the radial distance from the charged particle to the axis of its jet,
\begin{equation}
r=\sqrt{(\phi_{ch}-\phi_{jet})^2+(y_{ch}-y_{jet})^2},
\end{equation}
and the variable $p_T^{rel}$ refers to the momentum of charged particles in a jet, transverse to the jet axis,
\begin{equation}
p_T^{rel}=\frac{|\vec{p}_{ch}\times \vec{p}_{jet}|}{|\vec{p}_{jet}|^2},
\end{equation}
where $\vec{p}_{ch}$ and $\vec{p}_{jet}$
are the momentum of the charged particle and the jet, respectively.

\section {Forward method for jet study}
Beside the backward method we propose a forward method for the theoretical
study of the jet analysis. In the forward method, we first employ the
PYTHIA6.4 \cite{PYTHIA6.4} with the hadronization switched off temporarily to
generate the partonic initial state for the $pp$ collisions at $\sqrt{s}$=7
TeV. The partonic initial state is originally composed of simple (without gluons between
two quark ends) strings and complex (with gluons between two quark ends)
strings. The gluons are first removed from the complex strings. All the gluons are then
split into quark pairs, and each quark pair
is modeled as a simple string. Hence the partonic initial state of the $pp$ collision
is finally composed of simple strings only.

Secondly, each simple string in the the partonic initial state is hadronized
in the Lund string hadronization
model (i. e. by the subroutine of PYSTRF in the PYTHIA 6.4) individually. The
hadrons from the hadronization of a simple string are first cataloged into two
groups (related to the two quark ends) according to their relative transverse
momentum to the quark ends. The hadrons bearing momenta closer to the momentum of
one end of the string are grouped together. For instance, a hadron
is grouped as the candidate of the left quark end jet if its relative transverse
momentum to the left quark end is less than its relative transverse momentum
to the right quark end.

Thirdly, the candidate of a quark end jet is determined to be
the constituent of the jet if its distance relative to this
quark end satisfies Eqs. (1) and (2) ($R=0.6$) in the rapidity and azimuthal (y-$\phi$) phase space.
Finally, two original quark ends of a simple string are thus developed into two
hadronic jets, a two-to-two correspondence.

The routine in forward method is as follows: 1) We
create the partonic initial state by PYTHIA. 2) The simple
strings are then constructed. 3) Each simple string is
hadronized individually by Lund string fragmentation regime. 4)
Final hadrons from two quark ends in a simple string are
constructed to two jets. 5) The final hadronic state is
resulted eventually. The process is parallel to the natural time
evolution of collisions, as sketched in right part of Fig.(\ref{jet}).
Therefore it is referred to as forward method.

\section {Results}
To show the solidity of the forward method we first employ it to
calculate the final state charged particle transverse momentum and
pseudorapidity distributions in pp collisions at $\sqrt{s}$=7 TeV. The results
are shown in Figure (\ref{pt}) as triangles up and compared with the CMS
experimental data (solid squares) \cite{CMS}. In the calculation the K factor,
``multiplying the differential cross section for hard parton-parton processes" in
the PYTHIA6.4, is tuned to 4 from the default value of 1.5. One sees in Figure
(\ref{pt}) that both the theoretical transverse momentum distribution and
pseudorapidity distributions agree well with the CMS data. This tuned K factor of
4 is applied to all the later calculations.

\begin{widetext}
\begin{center}
\begin{figure}[htbp]
\includegraphics[width=0.45\textwidth]{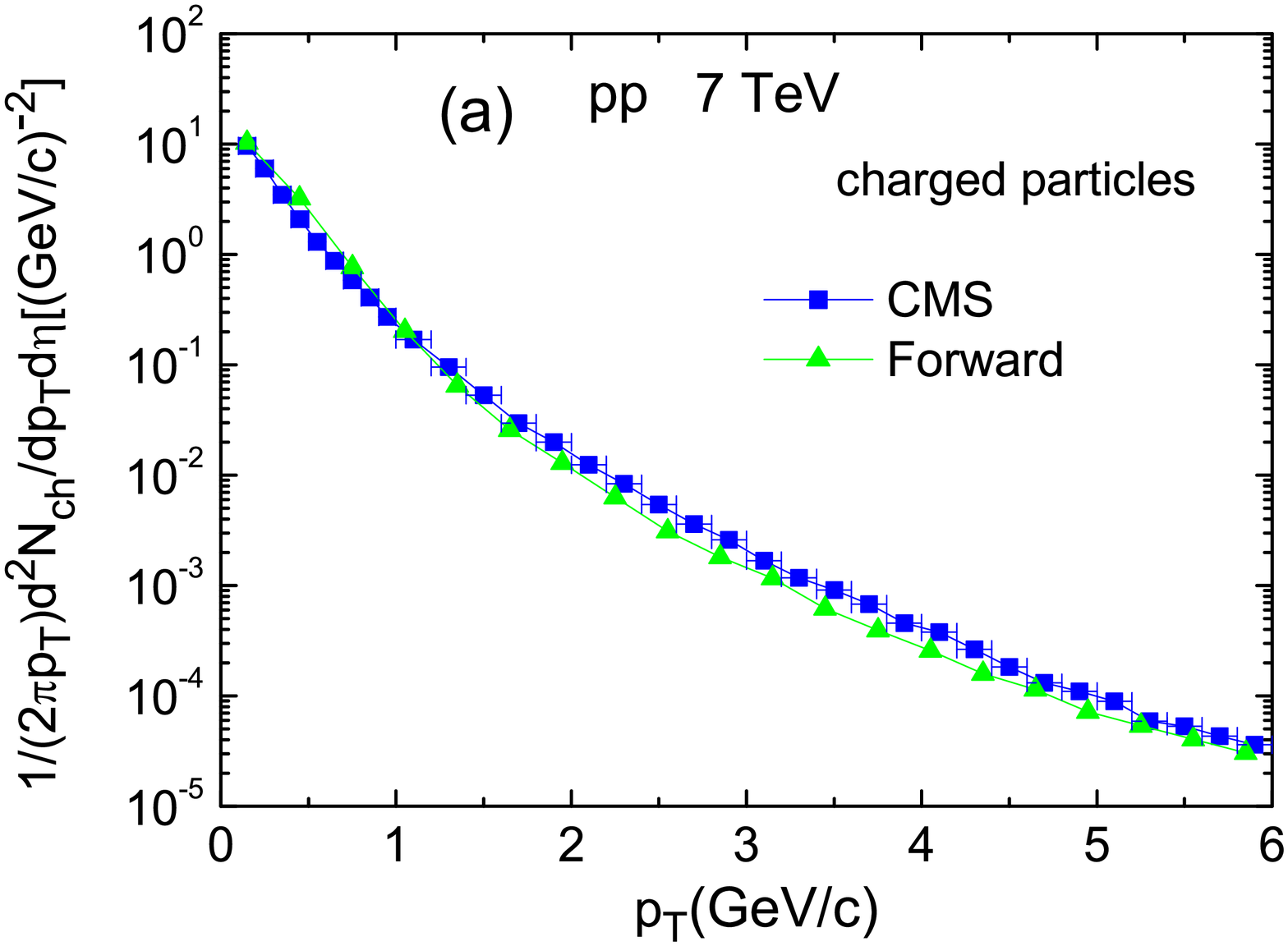}
\includegraphics[width=0.45\textwidth]{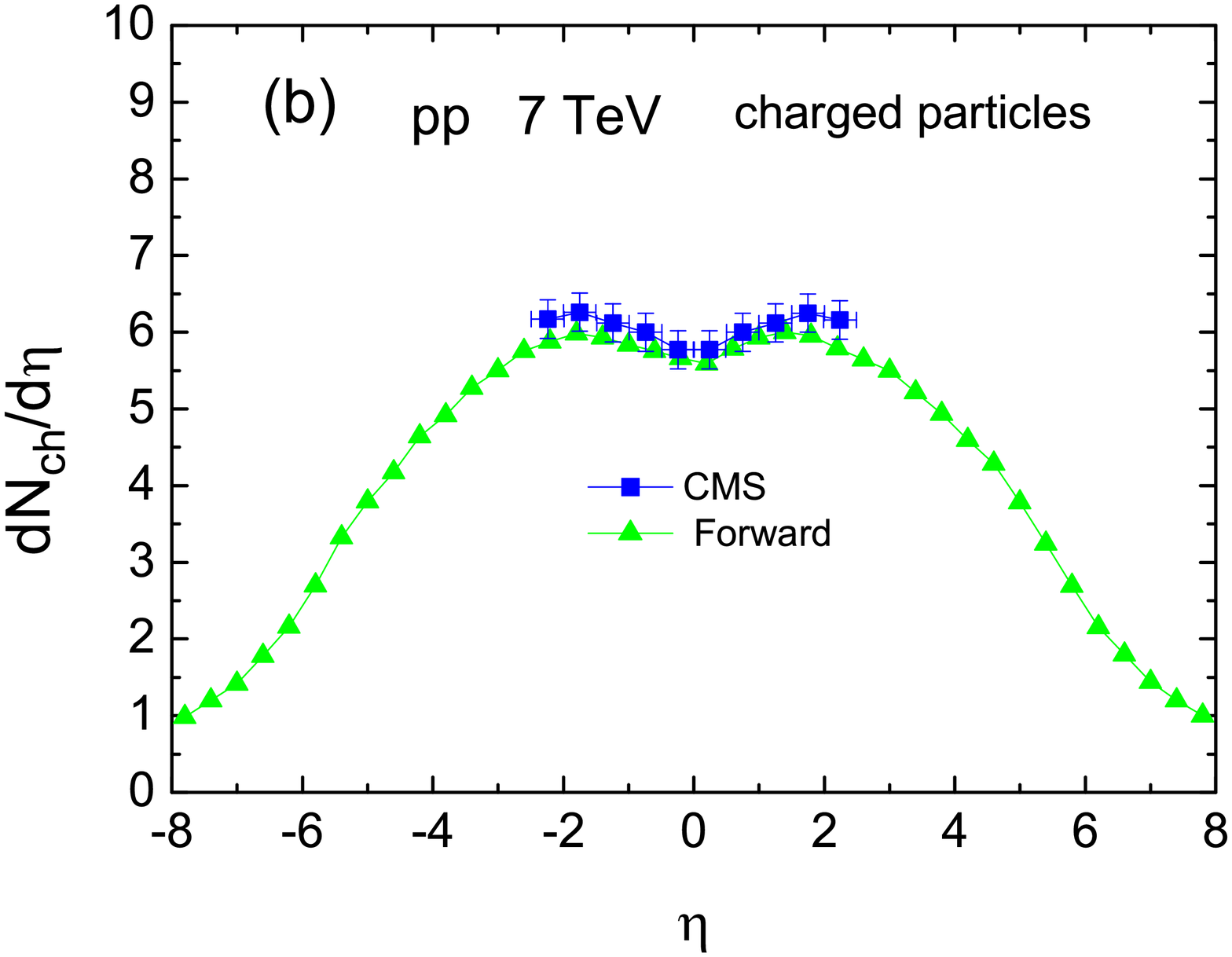}
\caption{(color online) The transverse momentum (a) and pseudorapidity (b)
distributions of final state charged particle in $pp$ collisions at $\sqrt{s}$=7
TeV calculated with forward method are compared with the CMS data (taken from
Ref. \cite{CMS})}.
\label{pt}
\end{figure}
\end{center}
\end{widetext}

Then the intra-jet charged particle distributions of $F(z, p_{\rm{T,jet}})$,
$f(p_T^{rel}, p_{\rm{T,jet}})$ and $\rho_{ch}(r,p_{\rm{T,jet}})$ calculated in the
backward method are compared with the corresponding ATLAS
experimental data \cite{ATLAS3} (analyzed by anti-$k_t$ algorithm, i. e. by
backward method) in Fig.(\ref{z}a)-(\ref{z}c) for the $pp$ collisions at $\sqrt{s}$
=7 TeV. We see in this figure that the theoretical results of the backward method
fairly well agree with the ATLAS data except the theoretical results of
$f(p_T^{rel}, p_{\rm{T,jet}})$ is smaller than the ATLAS data
at high $p_T^{rel}$ region for $p_{T,jet}$ bin of  10-15, 15-24 and 24-40.

\begin{widetext}
\begin{center}
\begin{figure}[htbp]
\includegraphics[width=0.9\textwidth]{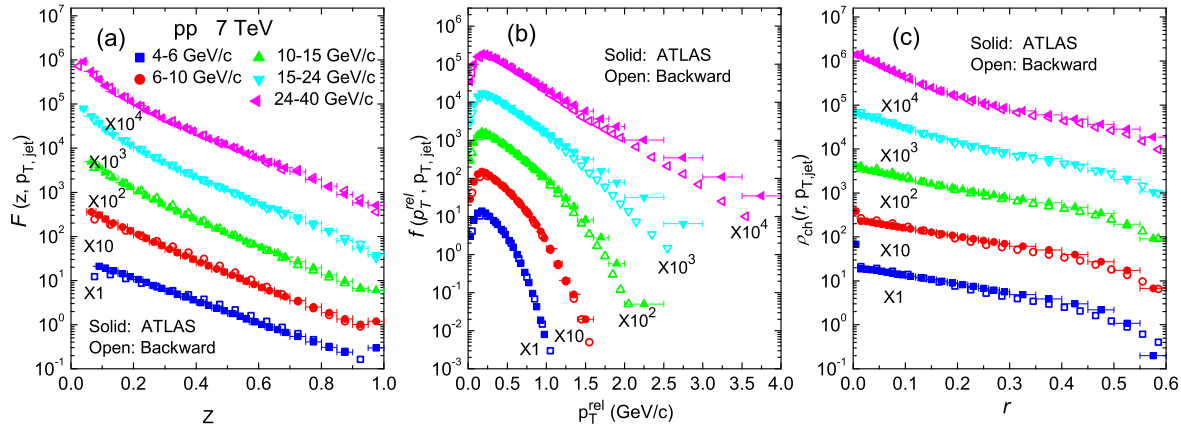}
\caption{(color online) The intra-jet charged particle distributions of (a)
$F(z, p_{\rm{T,jet}})$, (b) $f(p_T^{rel}, p_{\rm{T,jet}})$ and (c)
$\rho_{ch}(r, p_{\rm{T,jet}})$ in $p_{\rm{T,jet}}$ bins of 4-6, 6-10, 10-15, 15-24
and 24-40 GeV/c in $pp$ collisions at $\sqrt{s}$=7 TeV. The solid and open symbols
are the ATLAS data \cite{ATLAS3} and results of backward method, respectively. The
results in 6-10, 10-15, 15-24 and 24-40 GeV/c bins are multiplied by 10, $10^2$,
$10^3$ and $10^4$, respectively. The distance parameter $R$=0.6 is assumed.}
\label{z}
\end{figure}
\end{center}

\begin{center}
\begin{figure}[htbp]
\includegraphics[width=0.9\textwidth]{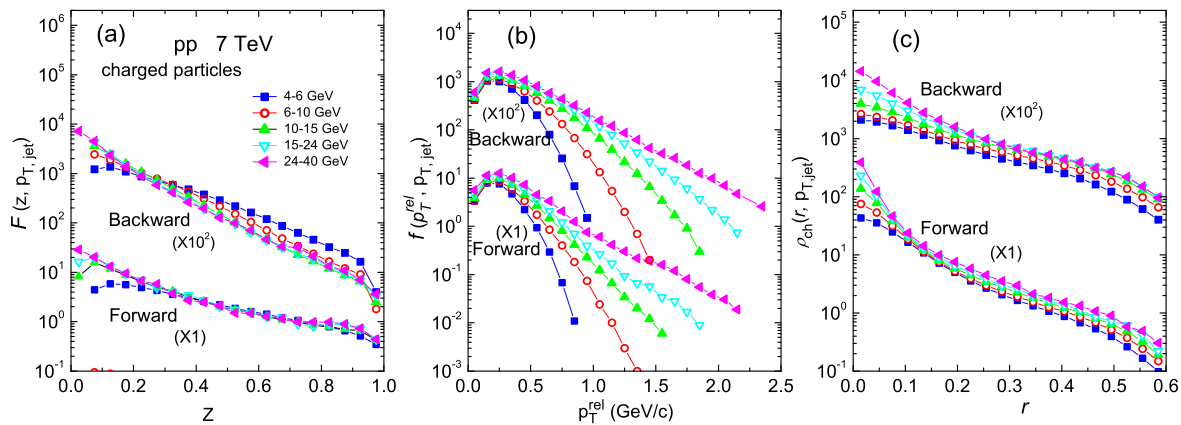}
\caption{(color online) The intra-jet charged particle distributions of (a)
$F(z, p_{\rm{T,jet}})$, (b) $f(p_T^{rel}, p_{\rm{T,jet}})$ and (c)
$\rho_{ch}(r, p_{\rm{T,jet}})$ calculated by forward method are compared to the
ones calculated by backward methods for $p_{\rm{T,jet}}$ bins of 4-6, 6-10, 10-15,
15-24 and 24-40 GeV/c in $pp$ collisions at $\sqrt{s}$=7 TeV. The results of
backward method are multiplied by $10^2$.}
\label{R2}
\end{figure}
\end{center}
\end{widetext}

Furthermore, the intra-jet charged particle distributions of $F(z, p_{\rm{T,jet}})$,
$f(p_T^{rel}, p_{\rm{T,jet}})$ and $\rho_{ch}(r,p_{\rm{T,jet}})$ calculated in the
forward method are compared with the ones calculated in the backward method
in Fig.(\ref{R2}a)-(\ref{R2}c) for $pp$ collisions at $\sqrt{s}$=7 TeV.
Fig. (\ref{R2}) shows that the results of the forward method are comparable with
the backward method ones. We see the scaling phenomena in the $F(z, p_{T,jet})$
distributions for the high $p_{T,jet}$ bin of 10-15, 15-24 and 24-40 in
backward methods and for all $p_{T,jet}$ bins in the forward method.

\section {Discussions and Conclusions}
In this paper a forward method based on PYTHIA6.4 is proposed to study
theoretically the properties of intra-jet charged particle distributions of
$F(z, p_{\rm{T,jet}})$, $f(p_T^{rel}, p_{\rm{T,jet}})$ and $\rho_{ch}(r,
p_{\rm{T,jet}})$, and the calculated results are compared with ones calculated in
backward method
(usual anti-$k_t$ algorithm). The ATLAS data (analyzed with anti-$k_t$ algorithm)
\cite{ATLAS3} of the intra-jet charged particle distributions of $F(z, p_{\rm{T,jet}})$
, $f(p_T^{rel}, p_{\rm{T,jet}})$ and $\rho_{ch}(r, p_{\rm{T,jet}})$ in
$pp$ collisions at $\sqrt{s}$=7 TeV are well reproduced in the backward method,
as shown in Fig.(\ref{z}). These distributions calculated in the forward method
are comparable with the ones calculated in the backward method, as shown in
Fig.(\ref{R2}a)-(\ref{R2}c).

By comparing the $F(z, p_{\rm{T,jet}})$, $f(p_T^{rel}, p_{\rm{T,jet}})$ and
$\rho_{ch}(r, p_{\rm{T,jet}})$ distributions resulted in the backward method to the
ones resulted in the forward method, some light might shed on the problem of
the correspondence between final hadronic jets and initial partons in the
backward method. The forward method is possible to relate the final hadronic
jet to its initial partonic origin, which helps to eliminate
the effect from combinatorial jets or fake jets. The results of
forward method can also serve a theoretical reference for the
results of backward method in jet study. Of course, it has to be further studied.

As mentioned in \cite{Connors}, one key problem of jet study in the backward method
both experimentally and theoretically is to identify the jets which are really
generated from hard scattered partons, and to eliminate the effect of
combinatorial jets and fake jets. However, this problem does not exist in the
forward method where the final hadronic jets are all constructed directly from the
initial partonic strings. Therefore, the comparison between the $F(z, p_{\rm{T,jet}})$,
$f(p_T^{rel}, p_{\rm{T,jet}})$ and $\rho_{ch}(r, p_{\rm{T,jet}})$ distributions in
the backward method and in the forward method
may help to evaluate and/or to eliminate the effect of combinatorial jets and fake jets in
the backward method. It needs to be further studied as well.

\textbf{Acknowledgments}
This work was supported by the National Natural Science
Foundation of China under grant Nos.: 11477130, 11775094 and by the
111 project of the foreign expert bureau of China. YLY, AL and YY
acknowledge the financial support from Suranaree University of
Technology and the Office of the Higher Education Commission under
the NRU project of Thailand. YLY acknowledge the financial support from
Key Laboratory of Quark and Lepton Physics in Central China Normal
University under grant No. QLPL201805 and the Continuous Basic
Scientific Research Project (No, WDJC-2019-13).

\end{document}